%% file: paper.tex
\documentclass[a4paper,10pt]{article}

\usepackage{geometry}
\usepackage[hidelinks]{hyperref} 

\usepackage{glossaries} 
\glsdisablehyper
\input{glossary}

\makeglossaries

\usepackage[T1]{fontenc}
\usepackage{upquote}

\usepackage{multirow}

\usepackage{url}

\usepackage{graphicx}

\usepackage{cite}

\begin{document}

\date{}

\title{Distributed Virtual Machine Consolidation: A Systematic \\ Mapping Study}

\author{Adnan Ashraf\footnote{Corresponding author e-mail: \texttt{adnan.ashraf@abo.fi}} , Benjamin Byholm, Ivan Porres \\
\small{Faculty of Natural Sciences and Technology} \\
\small{\AA bo Akademi University, Finland}
}

\maketitle

\begin{abstract}
\noindent
\textbf{Background:} \Gls{vm} consolidation is an effective technique to improve resource utilization and reduce energy footprint in cloud data centers. It can be implemented in a centralized or a distributed fashion. Distributed \gls{vm} consolidation approaches are currently gaining popularity because they are often more scalable than their centralized counterparts and they avoid a single point of failure.

\noindent
\textbf{Objective:} To present a comprehensive, unbiased overview of the state-of-the-art on distributed \gls{vm} consolidation approaches.

\noindent
\textbf{Method:} A \gls{sms} of the existing distributed \gls{vm} consolidation approaches.

\noindent
\textbf{Results:} 19 papers on distributed \gls{vm} consolidation categorized in a variety of ways. The results show that the existing distributed \gls{vm} consolidation approaches use four types of algorithms, optimize a number of different objectives, and are often evaluated with experiments involving simulations.

\noindent
\textbf{Conclusion:} There is currently an increasing amount of interest on developing and evaluating novel distributed \gls{vm} consolidation approaches. A number of research gaps exist where the focus of future research may be directed.

\vspace{6pt}
\noindent
\textbf{Keywords:} Cloud computing, Data center, Virtual machine, Consolidation, Placement, Energy-efficiency

\glsresetall
\end{abstract}

\section{Introduction}
\label{sec:background}
Energy footprint of cloud data centers is a matter of great concern for cloud providers~\cite{Mastelic:2014}. According to a recent report~\cite{Shehabi:2016}, data centers in the United States consumed an estimated 70 billion kilowatt-hours of electricity in 2014, which corresponds to 1.8\% of total United States electricity consumption. High energy consumption not only translates into a high operating cost, but also leads to huge carbon emissions. The ever increasing demand for computing resources to provide highly scalable and reliable services has caused an energy crisis~\cite{Kaur:2015:EET}. The high energy consumption of data centers can partly be attributed to the large-scale installations of computing and cooling infrastructures, but more importantly it is due to the inefficient use of the computing resources~\cite{Beloglazov:2012}. Production servers seldom operate near their full capacity~\cite{Barroso:2007}. However, even at the completely idle state, they consume a substantial proportion of their peak power~\cite{Fan:2007:PPW}. Therefore, under-utilized servers are highly inefficient.

Hardware virtualization technologies allow to share a \gls{pm} among multiple, performance-isolated platforms called \glspl{vm} to improve resource utilization. Further improvement in resource utilization and reduction in energy consumption can be achieved by consolidating \glspl{vm} on under-utilized \glspl{pm}. The basic idea of \gls{vm} consolidation is to migrate and place the \glspl{vm} on as few \glspl{pm} as possible and then release the remaining, unused \glspl{pm} for termination or for switching to a low-power mode to conserve energy. A \gls{vm} consolidation approach uses live \gls{vm} migration to consolidate \glspl{vm} on a reduced set of \glspl{pm}. \Gls{vm} consolidation has emerged as one of the most effective and promising techniques to reduce energy footprint of cloud data centers~\cite{Beloglazov:2012, Farahnakian:2015}.

\begin{figure}[t]
	\centering	
    \includegraphics[trim=10 50 40 10, clip, width=0.95\textwidth]{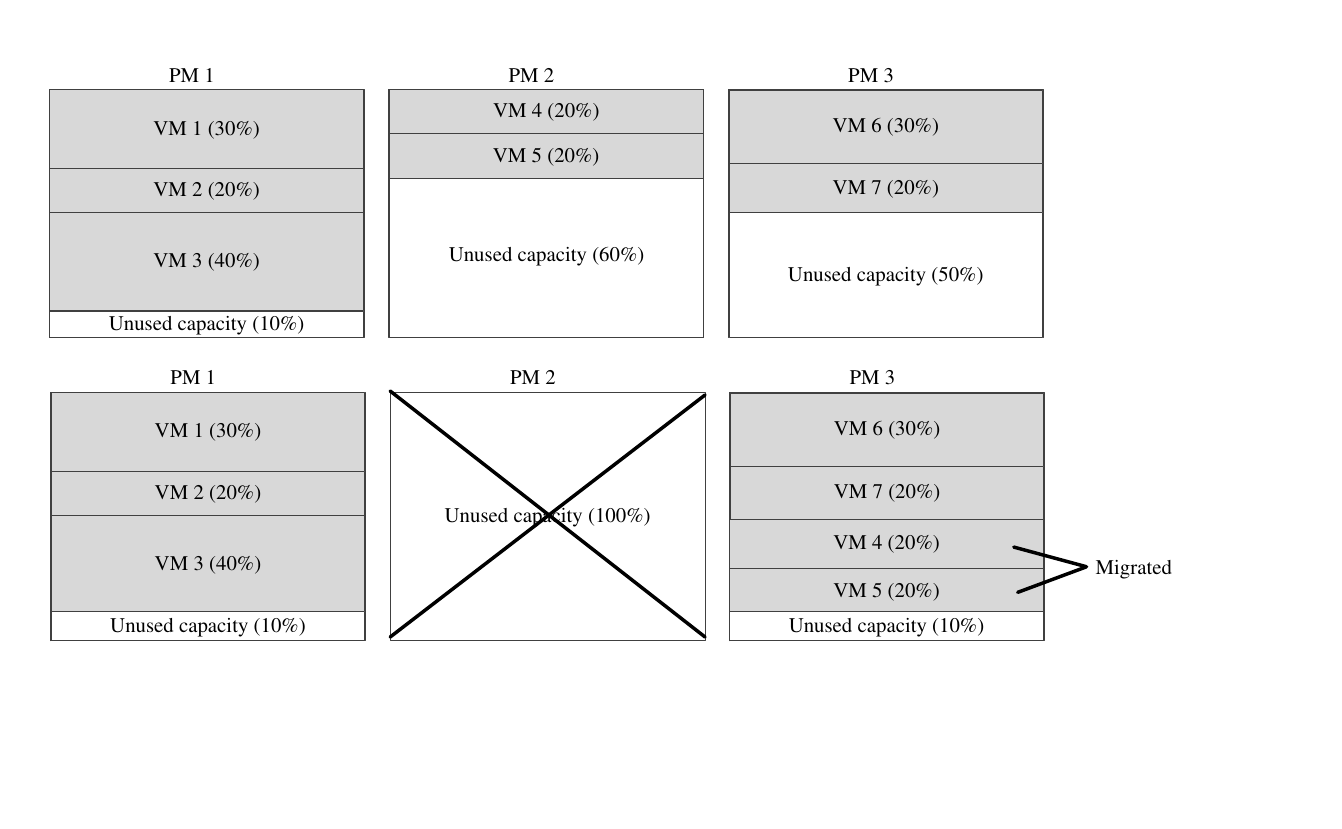}
	\caption{A simple example to illustrate the consolidation of \glspl{vm} on under-utilized \glspl{pm} to conserve energy}
	\label{fig:MotivatingExampleConsolidation}
\end{figure}

Figure~\ref{fig:MotivatingExampleConsolidation} presents a simple hypothetical scenario to illustrate the \gls{vm} consolidation process. The first half of Figure~\ref{fig:MotivatingExampleConsolidation} shows three \glspl{pm} where each \gls{pm} hosts multiple \glspl{vm} and every \gls{vm} uses a certain amount of the \gls{pm} resources. It is assumed that due to some significant load variations, \gls{pm} 2 and \gls{pm} 3 have become under-utilized. The under-utilized \glspl{pm} in such a scenario may continue to remain under-utilized for hours, days, or even weeks unless the existing \glspl{vm} require more resources or some new \glspl{vm} are placed on the under-utilized \glspl{pm}. Therefore, it is difficult to provide a resource and energy efficient allocation of \glspl{vm} without consolidation of \glspl{vm} on the under-utilized \glspl{pm}. The second half of Figure~\ref{fig:MotivatingExampleConsolidation} shows that after migrating all \glspl{vm} from \gls{pm} 2 to \gls{pm} 3, \gls{pm} 2 can be turned-off or switched to a low-power mode.

There is currently an increasing amount of interest on developing and evaluating efficient \gls{vm} consolidation approaches for cloud data centers. Over the past few years, researchers have used a multitude of ways to develop novel \gls{vm} consolidation approaches. Some of these approaches have been recently reported in the form of nonsystematic literature reviews such as~\cite{Ahmad:2015} and~\cite{Pires:2015}. However, the drawback of these existing nonsystematic studies is that they provide a partial and possibly biased overview of the state-of-the-art on \gls{vm} consolidation. For a comprehensive and unbiased coverage of the existing literature on \gls{vm} consolidation, there is a need to study the existing \gls{vm} consolidation approaches in a systematic way.

\Gls{vm} consolidation can be implemented in a centralized or a distributed fashion. Traditional \gls{vm} consolidation approaches, such as~\cite{Beloglazov:2012, Farahnakian2013UCC, Ferreto:2011, Hwang:2013, Ferdaus2014, Gao:2013, Farahnakian2014PDP}, tend to be centralized. A centralized \gls{vm} consolidation approach uses a centralized algorithm on a centralized architecture and does not provide support for multiple, geographically distributed data centers. The main drawbacks of centralized \gls{vm} consolidation approaches include limited scalability and lack of robustness due to a single point of failure. On the other hand, a distributed or decentralized \gls{vm} consolidation approach uses a distributed algorithm or a distributed architecture for \glspl{pm}~\cite{Farahnakian:2015, Feller2012CloudCom} or provides support for multiple, geographically distributed data centers~\cite{Lucanin:2015, Xu:2015:ICPADS}. Distributed \gls{vm} consolidation is a recurring theme in recent \gls{vm} consolidation approaches such as~\cite{Farahnakian:2015, Marzolla:2011, Feller2012CloudCom}. Distributed approaches are gaining popularity because they have benefits over centralized approaches. They are often more scalable than their centralized counterparts and they avoid a single point of failure~\cite{Marzolla:2011, Masoumzadeh:2015}. Feller et al.~\cite{Feller2012CloudCom} showed that their proposed \gls{vm} consolidation algorithm does not compute a solution (in a reasonable amount of time) on a centralized architecture, but finds a good solution on a distributed architecture. Lucanin and Brandic~\cite{Lucanin:2015} reported that their \gls{vm} consolidation algorithm for geographically distributed data centers finds a good solution for a large-scale problem comprising ten thousand \glspl{vm}. Sedaghat et al.~\cite{Sedaghat:2016} showed that their proposed distributed \gls{vm} consolidation algorithm scales to tens of thousands of \glspl{pm} and \glspl{vm} without compromising on the quality of the solution. Sedaghat et al.~\cite{Sedaghat:2014} reported that their proposed distributed \gls{vm} consolidation algorithm finds a near-optimal solution for 100,000 \glspl{pm} in a reasonable amount of time. Marzolla et al.~\cite{Marzolla:2011} showed that their proposed distributed \gls{vm} consolidation algorithm is resilient to major failures and outages involving a thousand \glspl{pm}. Therefore, distributed \gls{vm} consolidation approaches are more suitable for large-scale data centers involving thousands of \glspl{vm} and \glspl{pm}.

We present a systematic study of the existing distributed \gls{vm} consolidation approaches. The objective is to present a comprehensive, unbiased overview of the state-of-the-art on distributed \gls{vm} consolidation approaches. Considering the broad nature of the research objective, it was not appropriate to launch a \gls{slr}. Therefore, we launched a \gls{sms}~\cite{Petersen:2008, Budgen:2008, Petersen:2015}. A \gls{sms} follows the same principled process as a \gls{slr}, but: (1) it has a broader scope, (2) it uses different criteria for inclusions/exclusions and quality assessments, and (3) its data collection and synthesis tend to be more qualitative than for a \gls{slr}~\cite{Wohlin:2012}. It is \lq\lq intended to \lq map out\rq\ the research that has been undertaken rather than to answer a detailed research question\rq\rq~\cite{Budgen:2008}.

We proceed as follows. Section~\ref{sec:design} presents the design and schedule of our study. The results of the \gls{sms} are presented in Section~\ref{sec:results}. In Section~\ref{sec:validity}, we discuss major threats to the validity of the results presented in this paper. Finally, we present our conclusions in Section~\ref{sec:conclusions}.

\section{Study Design}
\label{sec:design}
One of the most important differences between a nonsystematic literature review and a \gls{sms} is that a \gls{sms} follows an unbiased and repeatable process. Moreover, the process is documented as a review protocol. Therefore, we defined a review protocol for our \gls{sms} on distributed \gls{vm} consolidation approaches. In this section, we present the design of our study and the review process.

\subsection{Research Questions}
\label{sec:rqs}
The \glspl{rq} are as follows:

\begin{itemize}
  \item \gls{rq}1: What approaches have been developed for distributed \gls{vm} consolidation?
  \item \gls{rq}2: What kinds of algorithms are being used in the existing distributed \gls{vm} consolidation approaches?
  \item \gls{rq}3: What objectives are being optimized in the existing distributed \gls{vm} consolidation approaches?
  \item \gls{rq}4: How are the existing distributed \gls{vm} consolidation approaches being evaluated?
  \item \gls{rq}5: What are the most popular publication forums for distributed \gls{vm} consolidation papers and how have they changed over time?
\end{itemize}

\Gls{rq}1 is the basic question for obtaining an overview of the state-of-the-art on distributed \gls{vm} consolidation. \Gls{rq}2 is aimed at obtaining the types of algorithms which are being used in the existing distributed \gls{vm} consolidation approaches. Possible types include heuristics, metaheuristics, and machine learning algorithms. Moreover, the algorithms may also be categorized into offline and online optimization algorithms.

\Gls{rq}3 concerns the objectives which are being optimized. Possible objectives include minimizing energy consumption, minimizing the number of active \glspl{pm}, minimizing \gls{sla} violations, minimizing the number of \gls{vm} migrations, minimizing cost, minimizing network traffic, maximizing performance, maximizing reliability, and minimizing resource utilization. \Gls{rq}3 also deals with the number of objectives which are being optimized and how the optimization problem is formulated. Possible problem formulations include single-objective, multi-objective (two or three objectives) with an \gls{aof}, pure multi-objective, and many-objective (four or more objectives)~\cite{Lucken:2014}.

\Gls{rq}4 concerns the evaluation method. The most common evaluation method in the \gls{vm} consolidation literature is experiment. An experiment may involve the use of prototype implementations or simulations. Moreover, the experiment design may involve realistic, synthetic, or hybrid load patterns. Similarly, an experimental evaluation may or may not include a comparison of the results with other existing \gls{vm} consolidation approaches. Finally, a comparison of the results may or may not include statistical tests to assess the statistical significance of the results.

\Gls{rq}5 is a typical question for \glspl{sms} in software engineering~\cite{Petersen:2008}. The objective is to identify the most popular, peer-reviewed publication forums with respect to distributed \gls{vm} consolidation papers. The publication forums may include journals, conferences, and workshops. In addition, the second part of \gls{rq}5 concerns the frequencies of published papers in popular forums over time to see the trends.

Based on the \glspl{rq}, the \gls{picoc}~\cite{Kitchenham:07} is presented in Table~\ref{tab:picoc}.

\begin{table}[!ht]
\centering
\small
\caption{\glsentrytext{picoc}}
\label{tab:picoc}
\begin{tabular}{|l|p{12.5cm}|}
\hline
\textbf{Aspect} & \textbf{Value}                                  \\ \hline
Population (P)      &  Energy-aware and cost-effective data center management techniques     \\ \hline
Intervention (I)      &  Distributed \gls{vm} consolidation/placement approaches/ algorithms/methods/heuristics  \\ \hline
Comparison (C)      &   No comparison intervention                                     \\ \hline
Outcomes (O)      &  An overview of the state-of-the-art on distributed \gls{vm} consolidation approaches       \\ \hline
Context (C)      & Cloud data centers            \\ \hline
\end{tabular}
\end{table}

\subsection{Search Strategy for Primary Studies}
\label{sec:search_strategy}
This section presents our search strategy. It is based on the \gls{slr} and \gls{sms} guidelines described by Kitchenham and Charters~\cite{Kitchenham:07} and Wohlin et al.~\cite{Wohlin:2012}.

\subsubsection{Search Terms}
Table~\ref{tab:terms} presents the most important search terms along with their alternate spellings. The search terms are primarily based on the \glspl{rq} and \gls{picoc} in Section~\ref{sec:rqs}. Moreover, they are also in line with recent and prominent works on \gls{vm} consolidation such as~\cite{Beloglazov:2012, Farahnakian:2015, Farahnakian2013UCC, Ferreto:2011, Hwang:2013, Marzolla:2011, Ferdaus2014, Gao:2013, Feller2012CloudCom, Farahnakian2014PDP}.

\begin{table}[!ht]
\centering
\small
\caption{Search terms}
\label{tab:terms}
\begin{tabular}{|l|l|p{7.2cm}|}
\hline
\textbf{\#} & \textbf{Search term} & \textbf{Alternate spellings}                          \\ \hline
1      & Consolidat* & Consolidate, consolidating, consolidation \\ \hline
2      & Plac* & Place, placing, placement                    \\ \hline
3      & Virtual machine* & Virtual machine, virtual machines \\ \hline
4      & \glsentrytext{vm}* & \glsentrytext{vm}, \glsentrytext{vm}s                                       \\ \hline
5      & Server* & Server, servers                \\ \hline
6      & Algorithm* & Algorithm, algorithms                \\ \hline
7      & Approach* & Approach, approaches                \\ \hline
8      & Method* & Method, methods                \\ \hline
9     & Heuristic* & Heuristic, heuristics                \\ \hline
10     & Cloud & None                \\ \hline
11     & Data center & Data center, datacenter, data centre, datacentre  \\ \hline
12     & Distributed & None \\ \hline
13     & Decentralized & None  \\ \hline
\end{tabular}
\end{table}

\subsubsection{Search Strings}
The search strings are presented in Table~\ref{tab:strings}. They are formed by making appropriate combinations of the search terms presented in Table~\ref{tab:terms}. We used Boolean \emph{AND} and Boolean \emph{OR} operators to make the search strings. The two search strings in Table~\ref{tab:strings} can also be combined into one search string by using the Boolean \emph{OR} operator. Therefore, the papers which contain any of the two search strings were retrieved. The search strings were validated against a set of known papers~\cite{Farahnakian:2015, Marzolla:2011, Feller2012CloudCom}.

\begin{table}[!ht]
\centering
\small
\caption{Search strings}
\label{tab:strings}
\begin{tabular}{|l|p{14.7cm}|}
\hline
\textbf{\#} & \textbf{Search string}                                  \\ \hline
1      & (Distributed OR Decentralized) AND Consolidat* AND ("virtual machine*" OR \glsentrytext{vm}* OR server*) AND (algorithm* OR approach* OR method* OR heuristic*) AND (cloud OR "data center" OR datacenter OR "data centre" OR datacentre) \\ \hline
2      & (Distributed OR Decentralized) AND Plac* AND ("virtual machine*" OR \glsentrytext{vm}*) AND (algorithm* OR approach* OR method* OR heuristic*) AND (cloud OR "data center" OR datacenter OR "data centre" OR datacentre)  \\ \hline
\end{tabular}
\end{table}

\subsubsection{Databases}

The search strings in Table~\ref{tab:strings} were searched in the publication title, abstract, and keywords. The following digital libraries were searched: (1) \gls{ieee} Xplore, (2) \gls{acm} Digital Library, (3) ScienceDirect, and (4) SpringerLink. The search strings were customized for each digital library. Moreover, since using multiple digital libraries creates duplicates, the search results were analyzed to identify and remove the duplicates.

\subsection{Study Selection Criteria}
\label{sec:selection_criteria}
This section presents our inclusion and exclusion criteria for primary studies.

\subsubsection{Inclusion Criteria}
The inclusion criteria for primary studies are as follows:
\begin{itemize}
  \item Distributed or decentralized \Gls{vm} consolidation approach or algorithm or method or heuristic \emph{AND}
  \item For cloud data center(s) \emph{AND}
  \item Written in English \emph{AND}
  \item Published in a peer-reviewed journal, conference, or workshop of computer science, computer engineering, or software engineering
\end{itemize}

In addition, if several papers presented the same \gls{vm} consolidation approach, only the most recent was included.

\subsubsection{Exclusion Criteria}

The exclusion criteria are the inverse of the inclusion criteria. If several papers presented the same \gls{vm} consolidation approach, all except the most recent were excluded.

\subsection{Study Selection Procedure}
\label{sec:selection_procedure}
The study selection procedure was applied on the search results to remove false positives. It comprises two phases, namely (1) title and abstract level screening and (2) full-text level screening.

\subsubsection{Title and Abstract Level Screening}
In this phase, the inclusion/exclusion criteria in Section~\ref{sec:selection_criteria} were applied to publication title and abstract. To minimize researcher bias, two researchers (first and second author) independently analyzed the search results. Afterwards, the results were compared and disagreements were resolved through discussions. Moreover, for any unresolved disagreements consensus meetings~\cite{Dyba:2008} involving all three researchers (first, second, and third author) were arranged. The short-listed studies from this phase were used as input for the second phase.

\subsubsection{Full-text Level Screening}
In this phase, the selected studies from the first phase were further analyzed on the basis of full-text. Two researchers (first and second author) independently applied the inclusion/exclusion criteria in Section~\ref{sec:selection_criteria} on the full-text. In this phase, the researchers also documented a reason for each excluded study~\cite{Usman:2014}. The results were compared in a similar way as in the first phase and disagreements between the researchers were resolved through discussions and consensus meetings. The output of this phase was the final short-listed set of primary studies.

\subsection{Study Quality Assessment Checklist and Procedure}
\label{sec:quality}
The final selected studies based on the selection procedure in Section~\ref{sec:selection_procedure} were assessed for their quality merit. To minimize researcher bias, two researchers (first and second author) independently assessed the quality merit of the final selected studies. Any studies not meeting the minimum quality requirements were excluded from the final set of primary studies.

Table~\ref{tab:checklist} presents the checklist for study quality assessment. For each question in the checklist, a three-level, numeric scale was used~\cite{Usman:2014}. The levels are: yes (2 points), no (0 point), and partial (1 point). Based on the checklist and the numeric scale, a study could score a maximum of $34$ and a minimum of $0$ points. Moreover, for each study, the two independent scores from the two researchers were aggregated by computing arithmetic mean. We used the first quartile ($34/4=8.5$) as the cutoff point for the inclusion of studies. Therefore, if a study scored less than $8.5$ points, it was excluded due to its lack of quality merit. The researchers documented the obtained score of each included/excluded study.

\begin{table}[!ht]
\centering
\small
\caption{Study quality assessment checklist}
\label{tab:checklist}
\begin{tabular}{|l|p{14.2cm}|}
\hline
\textbf{\#} & \textbf{Question}                                  \\ \hline
\multicolumn{2}{|l|}{\textbf{Theoretical contribution}}   \\ \hline
1      &  Is the \gls{vm} consolidation approach clearly described?    \\ \hline
2      &  Is the pseudocode of the proposed algorithm included and clearly described?   \\ \hline
3      &  Is a concise mathematical notation used?                       \\ \hline
4      &  Is the underlying theory clearly described?        \\ \hline
5      &  Are the assumptions clearly stated?        \\ \hline
6      &  Are the optimization objectives clearly stated?        \\ \hline
\multicolumn{2}{|l|}{\textbf{Experimental evaluation}}   \\ \hline
7      & Is the evaluation method clearly described?        \\ \hline
8      & Is a prototype implementation presented?            \\ \hline
9      & Is a simulation presented?            \\ \hline
10      & Is the experimental design clearly described?            \\ \hline
11      & Is the experimental setup clearly stated?            \\ \hline
12      & Are realistic load patterns used in the experimental design?    \\ \hline
13      & Are results from multiple different experiments included?           \\ \hline
14      & Are results from multiple runs of each experiment included?            \\ \hline
15      & Are the experimental results compared with other state-of-the-art \gls{vm} consolidation approaches?            \\ \hline
16      & Is a statistical test used to assess the statistical significance of the results?            \\ \hline
17      & Are the limitations or threats to validity clearly stated?            \\ \hline

\end{tabular}
\end{table}

\subsection{Data Extraction Strategy}
\label{sec:data_extraction}
Table~\ref{tab:data_extraction} presents the data extraction form which was used to record the data that the researchers extracted from the primary studies. Two researchers (first and second author) independently extracted data from all of the primary studies. The extracted data were used to classify the primary studies in a number of different ways. If a paper contained more than one \gls{vm} consolidation approaches, it was classified as coming under more than one heading~\cite{Budgen:2008}.

Budgen et al.~\cite{Budgen:2008} reported that differences in the classification of papers is a recurring theme in \glspl{sms}, even for experienced researchers. The two researchers differed considerably when classifying the papers according to the data extraction form in Table~\ref{tab:data_extraction}. The extracted data were compared and the differences were resolved in consensus meetings and by referring back to the original papers~\cite{Dyba:2008}.

\begin{table}[!ht]
\centering
\small
\caption{Data extraction form}
\label{tab:data_extraction}
\begin{tabular}{|p{12.5cm}|p{0.8cm}|p{1.6cm}|}
\hline
\textbf{Data item}                                                                                                              & \textbf{Value} & \textbf{Additional notes} \\ \hline
\multicolumn{3}{|l|}{\textbf{General}}                                                                                                                                       \\ \hline
Data extractor name                                                                                                             &                &                           \\ \hline
Data extraction date                                                                                                            &                &                           \\ \hline
Study \glsentrytext{id} (S1, S2, S3, ...)                                                                                           &                &                           \\ \hline
Bibliographic reference (title, authors, year, journal/conference/workshop name)                                                &                &                           \\ \hline
Author affiliations and countries                                                                                               &                &                           \\ \hline
Publication type (journal, conference, or workshop)                                                                             &                &                           \\ \hline
\multicolumn{3}{|l|}{\textbf{\glsentrytext{vm} consolidation related}}                                                                                                       \\ \hline
Type of algorithm used (e.g., heuristic, metaheuristic)                                                                         &                &                           \\ \hline
Specific algorithm used (e.g., first-fit decreasing, genetic algorithm)                                                         &                &                           \\ \hline
Online or offline optimization                                                                                                  &                &                           \\ \hline
Number of optimization objectives                                                                                               &                &                           \\ \hline
Name(s) of optimization objective(s)                                                                                            &                &                           \\ \hline
Problem formulation (e.g., single-objective, multi-objective with \glsentrytext{aof})                                           &                &                           \\ \hline
Name of evaluation method (e.g., analytical, experiment)                                                                        &                &                           \\ \hline
Evaluation tool (e.g., simulation, prototype implementation)                                                     &                &                           \\ \hline
Type of load pattern(s) used (realistic, synthetic, or hybrid)                                                                    &                &                           \\ \hline
Name(s) of other \gls{vm} consolidation approach(es) used for the comparison of the results                                           &                &                           \\ \hline
Study \glsentrytext{id}(s) of other \gls{vm} consolidation approach(es) used for the comparison of the results                               &                &                           \\ \hline
Name of the statistical test used (e.g., Wilcoxon Signed-Rank Test)                                                             &                &                           \\ \hline
Statistically significant results (yes, no)                                                                                     &                &                           \\ \hline
\end{tabular}
\end{table}

\subsection{Synthesis of the Extracted Data}
\label{sec:synthesis}
The extracted data based on the data extraction strategy in Section~\ref{sec:data_extraction} were synthesized separately for each \gls{rq}. The papers were categorized in a variety of dimensions and counts of the number of papers in different categories were recorded~\cite{Wohlin:2012}.

Perhaps the most important result of a \gls{sms} is a systematic map, which allows to identify evidence clusters and evidence deserts to direct the focus of future \glspl{slr} and to highlight areas where more primary studies should be performed~\cite{Kitchenham:07}. It is important to present the systematic map in an appropriate visual format that provides a quick overview of the field and supports better analyses~\cite{Petersen:2008}. Therefore, a visual representation of the systematic map was created by using appropriate chart types including a pie chart and several bubble charts~\cite{Petersen:2015}. The visual representation of the systematic map allowed thematic analysis~\cite{Wohlin:2012} to see which categories were well investigated and to identify research gaps~\cite{Petersen:2008}.

\subsection{Schedule of the Study}
\label{sec:timetable}
The initial draft of the review protocol was written on 09.3.2016. From 10.3.2016 to 18.3.2016, the protocol went through internal and external reviews resulting in several major and minor revisions. The internal reviewers were the authors themselves, while the external reviewer was Muhammad Usman\footnote{\url{https://www.bth.se/eng/staff/muhammad-usman-muu/}} from the Department of Software Engineering at Blekinge Institute of Technology.

Table~\ref{tab:timetable} presents the schedule of the \gls{sms}. The duration for title and abstract level screening was based on approximately 40 abstracts per day. Similarly, the duration for full-text level screening was based on approximately 5 full-text papers a day. The durations in Table~\ref{tab:timetable} implicitly also include approximately 10$\%$ time for consensus meetings to resolve disagreements among the researchers. It should be noted that the search for primary studies was completed on 24.3.2016. Therefore, any papers published after this date are not covered in the mapping study.

\begin{table}[!ht]
\centering
\small
\caption{Schedule of the study}
\label{tab:timetable}
\begin{tabular}{|l|l|l|l|}
\hline
\textbf{\#} & \textbf{Activity}                  & \textbf{Duration} & \textbf{Due date} \\ \hline
1           & Protocol review and revision       & 07 business days            & 18.3.2016         \\ \hline
2           & Search for primary studies         & 04 business days            & 24.3.2016         \\ \hline
3           & Title and abstract level screening & 07 business days          & 06.4.2016          \\ \hline
4           & Full-text level screening          & 07 business days         & 15.4.2016         \\ \hline
5           & Quality assessment                 & 02 business days            & 19.4.2016          \\ \hline
6           & Data extraction                    & 05 business days           & 26.4.2016         \\ \hline
7           & Data synthesis                     & 05 business days           & 03.5.2016         \\ \hline
8           & Initial draft of paper              & 10 business days            & 18.5.2016          \\ \hline
\end{tabular}
\end{table}

\section{Results}
\label{sec:results}
In this section, we present the results of the \gls{sms} on distributed \gls{vm} consolidation approaches. Table~\ref{tab:paper_count_stages} presents the number of papers in different stages of the \gls{sms}. The results show that the initial search retrieved 202 results. However, out of 202, 86 were found duplicate and were subsequently removed. The main reason for such a large number of duplicate results is that the results from the \gls{acm} Digital Library were based on the \gls{acm} Guide to Computing Literature, which provides an expanded search that includes papers from the \gls{acm} full-text collection as well as from a number of other digital libraries including \gls{ieee} Xplore, ScienceDirect, and SpringerLink. The advantage of using the \gls{acm} Guide to Computing Literature is that it often finds more papers, but the disadvantage is that some of those papers are also found in the \gls{ieee} Xplore, ScienceDirect, and SpringerLink digital libraries. There were also a few cases where a \gls{vm} consolidation approach was published in several papers. In such cases, only the most recent paper was included.

\begin{table}[!ht]
\centering
\small
\caption{Number of papers in different stages}
\label{tab:paper_count_stages}
\begin{tabular}{|l|c|}
\hline
\textbf{SMS Stage}                       & \textbf{Number of papers} \\ \hline
Initial search results                   & 202                       \\ \hline
After removing duplicates                & 116                       \\ \hline
After title and abstract level screening & 41                        \\ \hline
After full-text level screening          & 21                        \\ \hline
After quality assessment                 & 19                        \\ \hline
\end{tabular}
\end{table}

From the remaining 116 papers, 75 papers were removed in the title and abstract level screening, resulting in 41 papers. Another 20 papers were removed in the full-text level screening. Out of these, 15 did not present a \gls{vm} consolidation approach~\cite{Simonin:2013, Wang:2009, Elmroth:2011, Jing:2015, Kabir:2014, Khosravi:2013, Pham:2016:HVM, Teyeb:2014, Jung:2010, Ravi:2014, Wen:2011, Abdul:2014, Chen:2013:ICC, Gu:2015:ICNC, Islam:2015}, 2 presented a centralized \gls{vm} consolidation approach~\cite{Feller:2011, Hermenier:2011}, 2 were published in several papers~\cite{Gutierrez:2013, Masoumzadeh:2013}, and 1 had a duplicate record~\cite{Sedaghat:2016}. Finally, from the remaining 21 papers, 2 papers were removed in the quality assessment stage because their aggregate scores were below the cutoff point~\cite{Larsson:2011:SMI, Donadio:2014}. It resulted in a total of 19 papers for data extraction and synthesis. Table~\ref{tab:final_papers} presents study \glspl{id} and references of the final selected primary studies. Each primary study in Table~\ref{tab:final_papers} presents a distributed \gls{vm} consolidation approach.

\begin{table}[!ht]
\centering
\scriptsize 
\caption{Study \glsentrytext{id}s and references of the final selected primary studies}
\label{tab:final_papers}
\begin{tabular}{|l|p{2.6cm}|p{10.0cm}|l|l|}
\hline
\textbf{\glsentrytext{id}} & \textbf{Authors} & \textbf{Title} & \textbf{Year} & \textbf{Ref.} \\ \hline
S1 & Sedaghat et al. & Decentralized cloud datacenter reconsolidation through emergent and topology-aware behavior & 2016 & \cite{Sedaghat:2016} \\ \hline
S2 & Zhang et al. & Dynamic service placement in shared service hosting infrastructures & 2010 & \cite{Zhang:2010:DSP} \\ \hline
S3 & Farahnakian et al. & Using ant colony system to consolidate VMs for green cloud computing & 2015 & \cite{Farahnakian:2015} \\ \hline
S4 & Feller et al. & A case for fully decentralized dynamic VM consolidation in clouds & 2012 & \cite{Feller2012CloudCom} \\ \hline
S5 & Gutierrez-Garcia and Ramirez-Nafarrate & Collaborative agents for distributed load management in cloud data centers using live migration of virtual machines & 2015 & \cite{Gutierrez:2015} \\ \hline
S6 & Lucanin and Brandic & Pervasive cloud controller for geotemporal inputs & 2015 & \cite{Lucanin:2015} \\ \hline
S7 & Marzolla et al. & Server consolidation in clouds through gossiping & 2011 & \cite{Marzolla:2011} \\ \hline
S8 & Masoumzadeh and Hlavacs & A cooperative multi agent learning approach to manage physical host nodes for dynamic consolidation of virtual machines & 2015 & \cite{Masoumzadeh:2015} \\ \hline
S9 & Mastroianni et al. & Analysis of a self-organizing algorithm for energy saving in data centers & 2013 & \cite{Mastroianni:2013} \\ \hline
S10 & Pham et al. & Joint consolidation and service-aware load balancing for datacenters & 2016 & \cite{Pham:2016:CL} \\ \hline
S11 & Sedaghat et al. & Divide the task, multiply the outcome: cooperative VM consolidation & 2014 & \cite{Sedaghat:2014} \\ \hline
S12 & Wang et al. & Multiagent-based resource allocation for energy minimization in cloud computing systems & 2016 & \cite{Wang:2016} \\ \hline
S13 & Xu et al. & Electricity cost minimization in distributed clouds by exploring heterogeneity of cloud resources and user demands & 2015 & \cite{Xu:2015:ICPADS} \\ \hline
S14 & Adhikary et al. & Energy-efficient scheduling algorithms for data center resources in cloud computing & 2013 & \cite{Adhikary:2013} \\ \hline
S15 & Mehta et al. & Energy cost management for geographically distributed data centres under time-variable demands and energy prices & 2013 & \cite{Mehta:2013} \\ \hline
S16 & Sakumoto et al. & Autonomous decentralized mechanisms for generating global order in large-scale system: using Metropolis-Hastings algorithm and applying to virtual machine placement & 2012 & \cite{Sakumoto:2012} \\ \hline
S17 & Sonnek et al. & Starling: minimizing communication overhead in virtualized computing platforms using decentralized affinity-aware migration & 2010 & \cite{Sonnek:2010} \\ \hline
S18 & Velasco et al. & Elastic operations in federated datacenters for performance and cost optimization & 2014 & \cite{Velasco:2014} \\ \hline
S19 & Kavvadia et al. & Elastic virtual machine placement in cloud computing network environments & 2015 & \cite{Kavvadia:2015} \\ \hline
\end{tabular}
\end{table}

\begin{figure}[!hb]
	\centering	
    \includegraphics[trim=140 140 210 150, clip, width=0.79\textwidth]{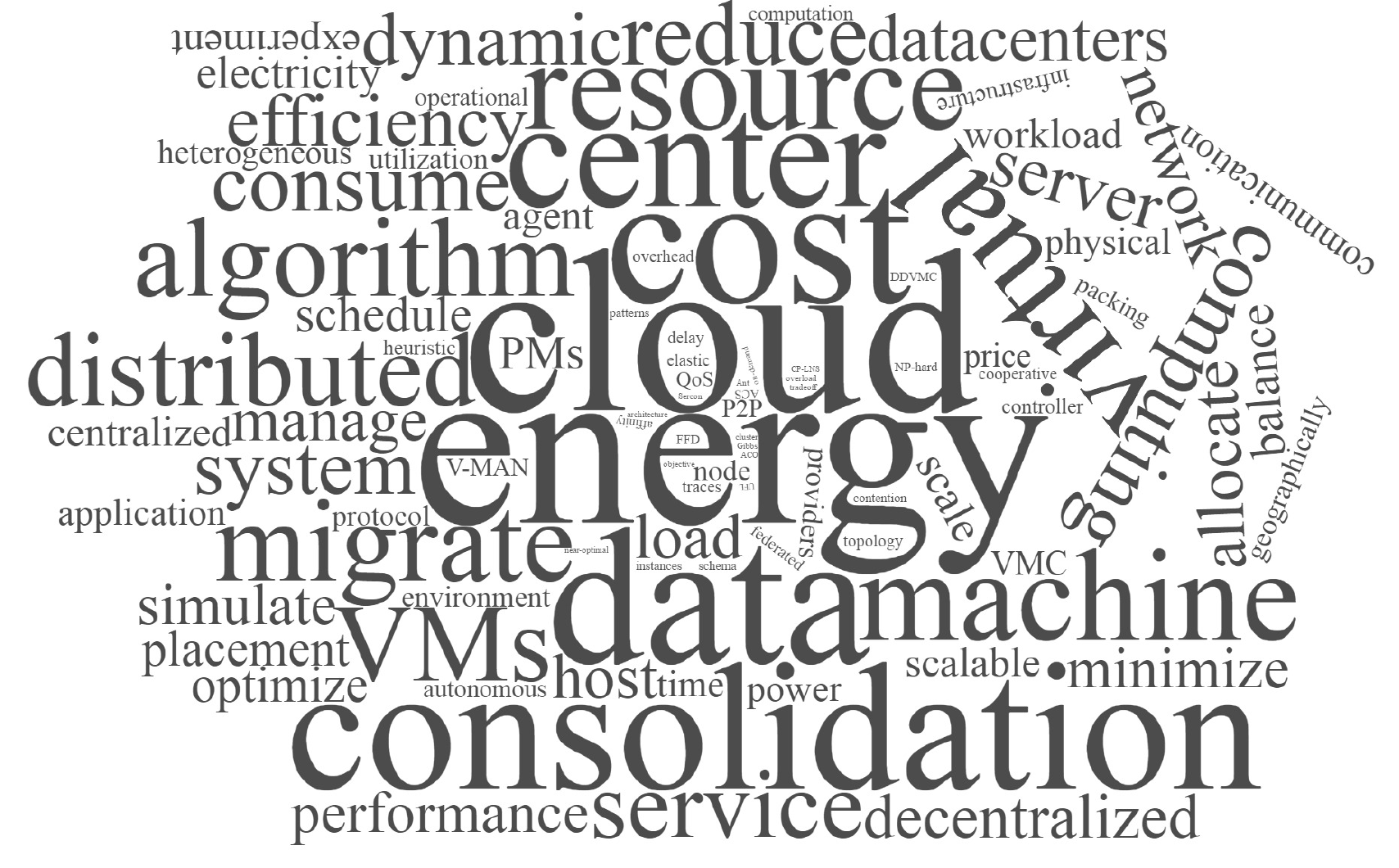}
	\caption{Word cloud of the titles and abstracts of the 19 primary studies}
	\label{fig:wordcloud}
\end{figure}

Before presenting detailed results, we start with a simple, graphical overview of the topic. Figure~\ref{fig:wordcloud} presents a word cloud of distributed \gls{vm} consolidation approaches generated from the titles and abstracts of the 19 primary studies. Common English words and those appearing only once were removed from the word list. Moreover, different forms and alternate spellings of a word were aggregated. The word cloud shows that some of the most frequent words include energy, cloud, data, and cost, which appear 56, 55, 46, and 45 times, respectively.

\subsection{RQ1: Distributed VM Consolidation Approaches}
The 19 distributed \gls{vm} consolidation approaches can be classified into three distinct categories: (1) pure distributed \gls{vm} consolidation algorithms, (2) centralized algorithms with a distributed architecture for \gls{vm} consolidation, and (3) \gls{vm} consolidation algorithms for geographically distributed data centers. Table~\ref{tab:three_categories} presents the three categories of distributed \gls{vm} consolidation approaches. A pure distributed \gls{vm} consolidation approach uses a distributed rather than a centralized algorithm to find a migration plan that optimizes the placement of \glspl{vm} on \glspl{pm}. In contrast, a distributed architecture approach provides a decentralized schema or multi-agent architecture for \glspl{pm}~\cite{Farahnakian:2015, Feller2012CloudCom}, but implements a centralized \gls{vm} consolidation algorithm. The approaches in the third category extend centralized algorithms to support \gls{vm} consolidation in multiple, geographically distributed data centers~\cite{Lucanin:2015, Xu:2015:ICPADS}. Table~\ref{tab:three_categories} shows that 14 primary studies present pure distributed \gls{vm} consolidation algorithms, 2 present centralized algorithms with a distributed architecture, and 3 present \gls{vm} consolidation algorithms for distributed data centers.

\begin{table}[!ht]
\centering
\small
\caption{Categories of distributed \glsentrytext{vm} consolidation approaches}
\label{tab:three_categories}
\begin{tabular}{|l|p{8.9cm}|c|}
\hline
\textbf{Category}                           & \textbf{Study \glsentrytext{id}s and references} & \textbf{Count} \\ \hline
Pure distributed                            & S1, S2, S5, S7, S8, S9, S10, S11, S12, S14, S16, S17, S18, S19      & 14               \\ \hline
Distributed architecture                    & S3, S4                                                              &  2             \\ \hline
For distributed data centers                & S6, S13, S15                                                        &  3              \\ \hline
\end{tabular}
\end{table}

\subsection{RQ2: Algorithm Types and Names}
Table~\ref{tab:synthesis_1} presents algorithm types and algorithm names along with the study \glspl{id} and counts. The results show that the most frequently used algorithm in distributed \gls{vm} consolidation approaches is distributed or coordinated local search heuristic, which was used in six primary studies. The second most used algorithms are \gls{aco} and greedy, which were both used in two primary studies. All other algorithms were used in only one primary study. The results also show that S6 appears under two headings, namely greedy and genetic algorithm. This is because S6 presents a hybrid, two-stage optimization approach that combines a genetic algorithm with a greedy, best-fit approach for local improvement.

\begin{table}[!ht]
\centering
\small
\caption{Types of algorithms and specific algorithms}
\label{tab:synthesis_1}
\begin{tabular}{|p{2.5cm}|p{7.5cm}|p{3.5cm}|c|}
\hline
\textbf{Type} & \textbf{Algorithm name} & \textbf{Study \glsentrytext{id}s} & \textbf{Count} \\ \hline
\multirow{5}{*}{Heuristic}	& Distributed or coordinated local search	& S1, S5, S7, S11, S12, S19	& 6 \\ \cline{2-4}
	& Greedy	& S6, S18	& 2 \\ \cline{2-4}
	& Weighted maximum bipartite matching	& S13	& 1 \\ \cline{2-4}
	& Static threshold-based	& S14	& 1 \\ \cline{2-4}
	& Distributed bartering algorithm	& S17	& 1 \\ \hline
\multirow{4}{*}{Metaheuristic}	& Local search	& S2	& 1 \\ \cline{2-4}
	& \glsentrytext{aco}	& S3, S4	& 2 \\ \cline{2-4}
	& Genetic algorithm	& S6	& 1 \\ \cline{2-4}
	& Constraint programming-based large neighborhood search	& S15	& 1 \\ \hline
Machine learning	& Fuzzy Q-learning algorithm	& S8	& 1 \\ \hline
\multirow{3}{*}{Statistical}	& Probabilistic search	& S9	& 1 \\ \cline{2-4}
	& Gibbs sampling and the alternating direction method of multipliers	& S10	& 1 \\ \cline{2-4}
	& Metropolis-Hastings algorithm	& S16	& 1 \\ \hline
\end{tabular}
\end{table}

\begin{figure}[!ht]
	\centering	
    \includegraphics[width=0.69\textwidth]{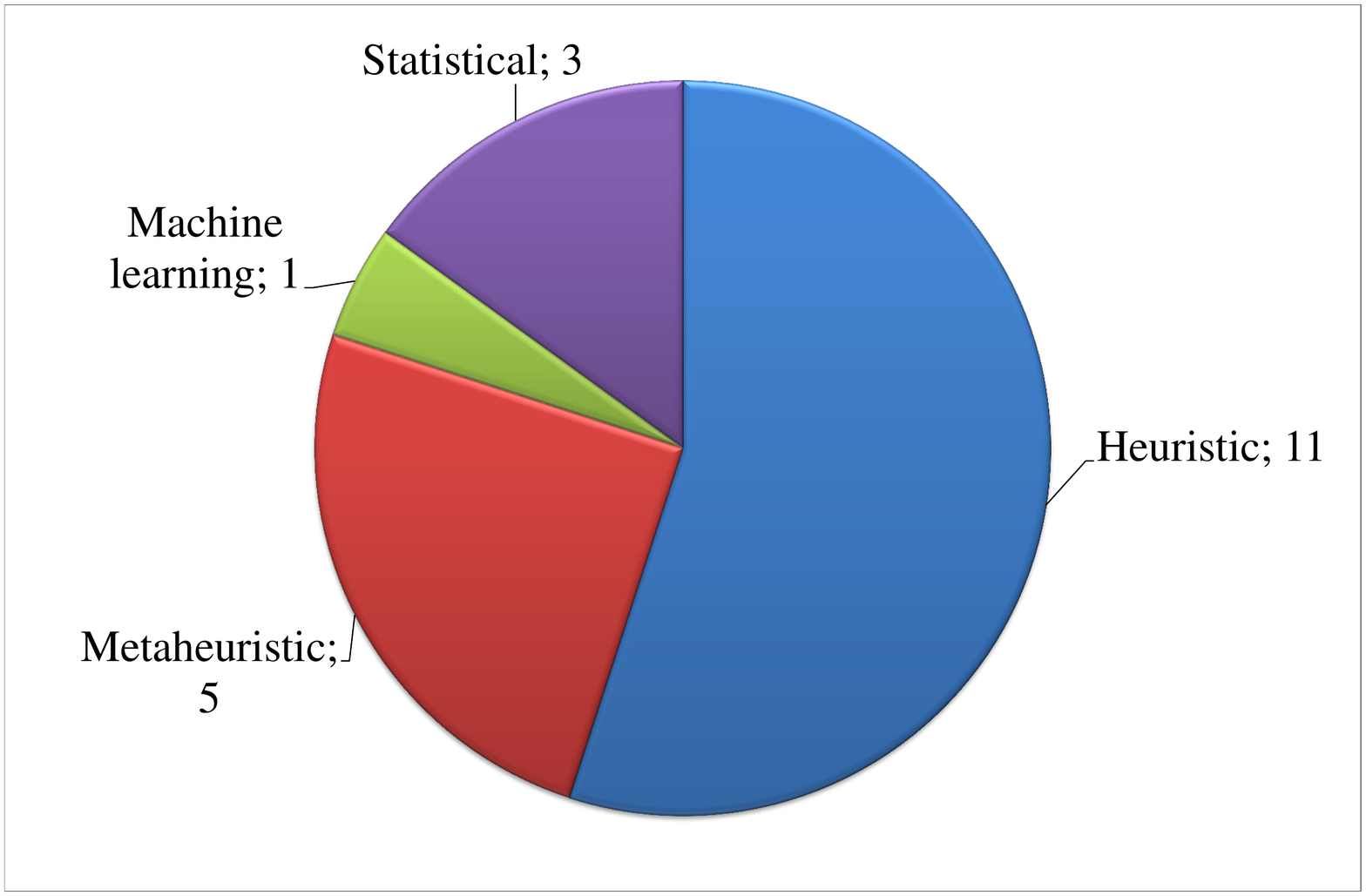}
	\caption{Algorithm types}
	\label{fig:algo_types}
\end{figure}

Figure~\ref{fig:algo_types} presents a summary of the four algorithm types from Table~\ref{tab:synthesis_1}. It shows that 11 primary studies presented a heuristic approach, 5 presented a metaheuristic, 3 presented a statistical approach, and 1 presented a machine learning approach. Therefore, the results indicate that heuristics and metaheuristics are currently the most popular algorithm types for distributed \gls{vm} consolidation, which collectively account for 80\% of all algorithm types.

Table~\ref{tab:synthesis_2} presents a classification of the primary studies with respect to online and offline optimization techniques. The results show that only 2 primary studies presented an online optimization approach, while 14 primary studies are based on offline optimization. Moreover, 3 primary studies were classified as inconclusive because the papers do not contain sufficient information in this regard. Therefore, the results indicate that 89\% of the primary studies either use offline optimization or are classified as inconclusive. This is an important result and is in line with past research in this area. Other researchers have also found that most of the optimization problems in cloud computing are currently being addressed with offline optimization techniques~\cite{Harman2013}. However, a drawback of the offline optimization techniques is that they require complete knowledge of the problem including possible future events, which is often difficult in a real world setting. In contrast, an online optimization algorithm receives input and produces output in an online manner~\cite{Beloglazov:2012}. Therefore, to be able to better cope with workload variability of different types of modern applications, \gls{vm} consolidation should be performed continuously in an online manner~\cite{Beloglazov:2012, Farahnakian:2015}. Hence, there is a clear need to develop online optimization approaches for distributed \gls{vm} consolidation.

\begin{table}[!ht]
\centering
\small
\caption{Online or offline optimization}
\label{tab:synthesis_2}
\begin{tabular}{|l|p{9.0cm}|c|}
\hline
\textbf{Online or offline} & \textbf{Study \glsentrytext{id}s} & \textbf{Count} \\ \hline
Online	& S3, S9 & 2 \\ \hline
Offline	& S2, S4, S5, S6, S7, S10, S12, S13, S14, S15, S16, S17, S18, S19 & 14 \\ \hline
Inconclusive	& S1, S8, S11 & 3 \\ \hline
\end{tabular}
\end{table}

\subsection{RQ3: Objectives}
Table~\ref{tab:synthesis_4} categorizes the primary studies with respect to the different problem formulations and the number of optimization objectives. The results show that 11 studies proposed a multi-objective (two or three objectives) problem formulation with an \gls{aof}. Similarly, 2 studies presented a many-objective (four or more objectives)~\cite{Lucken:2014} problem formulation with an \gls{aof}. The remaining 6 studies proposed a single-objective problem formulation. The results also indicate that 74\% of the primary studies optimize either one or two objectives. Hence, there is currently more published work on approaches that optimize fewer objectives. It should be noted that, although 13 approaches optimize two or more objectives, none of the studies proposed a pure multi-objective or many-objective problem formulation. The benefit of the \gls{aof} approach is that it reduces complexity and often improves runtime of the algorithm by limiting the search to a subspace of the feasible solutions. However, the drawback is that a correct combination of the objectives requires certain weights to be assigned to each objective, which often requires an in-depth knowledge of the problem domain~\cite{Pires:2015}. Therefore, the assignment of the weights is essentially subjective~\cite{Hu:2013:Pareto}. In contrast, pure multi-objective and many-objective approaches do not require weights and often provide a more elaborate search of the search space. Hence, there exists a research gap. There is a need to develop novel distributed \gls{vm} consolidation approaches that formulate the problem as pure multi-objective or many-objective.

\begin{table}[!ht]
\centering
\small
\caption{Problem formulations and number of optimization objectives}
\label{tab:synthesis_4}
\begin{tabular}{|p{3.9cm}|l|p{4.8cm}|c|}
\hline
\textbf{Problem formulation}                                       & \textbf{Objectives} & \textbf{Study \glsentrytext{id}s} & \textbf{Count} \\ \hline
Single-objective                                                   & 1 objective                   & S7, S12, S13, S14, S17, S19         & 6              \\ \hline
\multirow{2}{3.9cm}{Multi-objective with \glsentrytext{aof}}       & 2 objectives                  & S2, S4, S6, S8, S9, S10, S11, S16   & 8              \\ \cline{2-4}
                                                                   & 3 objectives                  & S3, S5, S18                         & 3              \\ \hline
\multirow{2}{3.9cm}{Many-objective with \glsentrytext{aof}}        & 4 objectives                  & S15                                 & 1              \\ \cline{2-4}
                                                                   & 5 objectives                  & S1                                  & 1              \\ \hline
\end{tabular}
\end{table}

\begin{figure}[!ht]
	\centering	
    \includegraphics[width=0.79\textwidth]{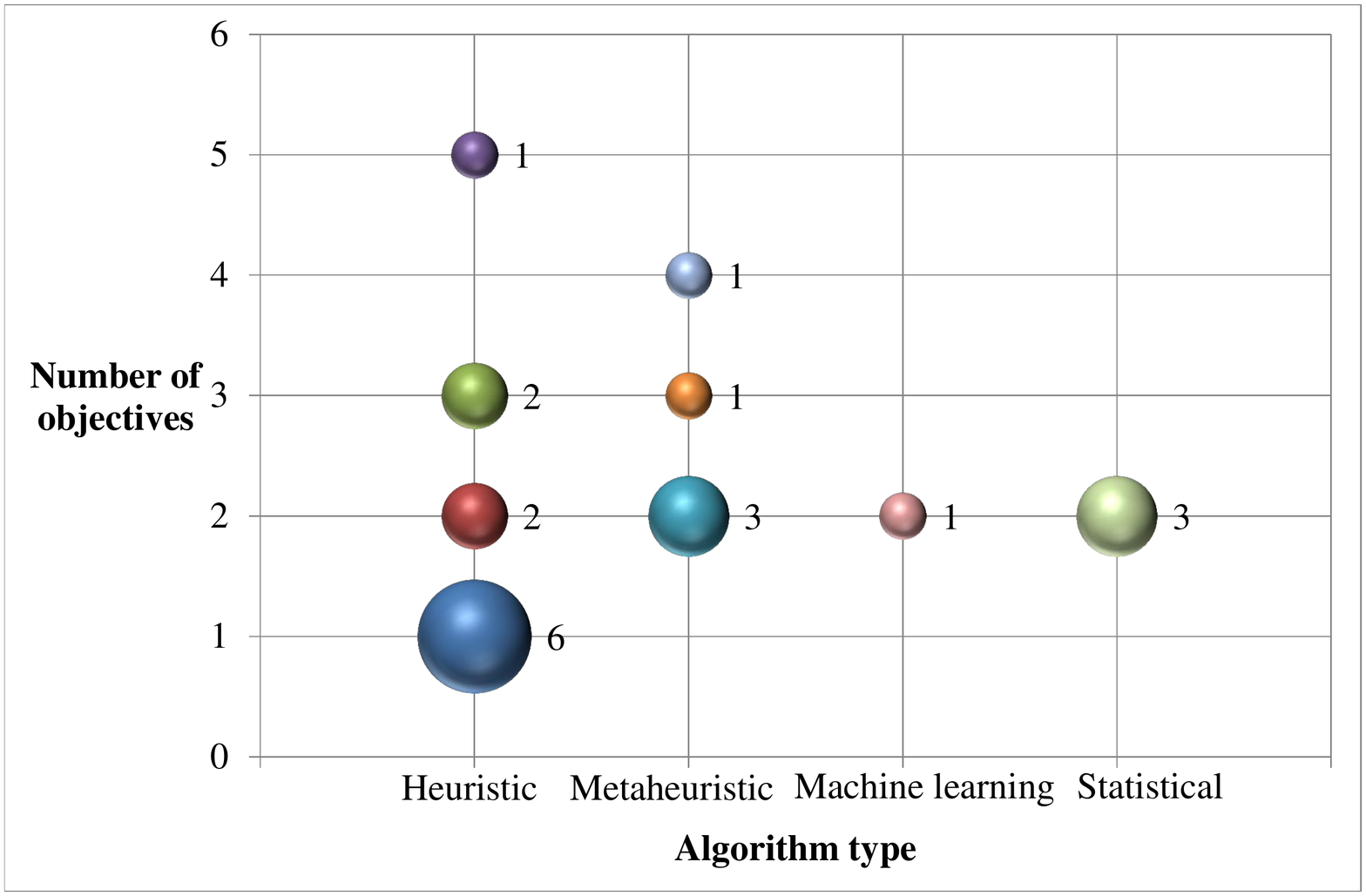}
	\caption{Number of primary studies (as bubble size) with respect to algorithm types and number of optimization objectives}
	\label{fig:algo_types_objectives}
\end{figure}

Figure~\ref{fig:algo_types_objectives} presents the number of primary studies (as bubble size) with respect to the four different algorithm types and the five different number of objectives. Again, S6 appears twice. The results show that in 6 out of 11 studies involving heuristics, a simpler problem formulation was proposed because it optimizes a single objective. Whereas, none of the other algorithm types was used for single-objective optimization. Therefore, although heuristics are currently the most published algorithm type for distributed \gls{vm} consolidation, they are mostly used for simpler, single-objective problem formulations.

Table~\ref{tab:synthesis_5} categorizes the primary studies on the basis of different optimization objectives. The results show that minimizing energy consumption is the most commonly found optimization objective in the currently published distributed \gls{vm} consolidation approaches as it appears in 11 out of 19 studies. Other popular optimization objectives include minimizing \gls{sla} violations and minimizing the number of \gls{vm} migrations, which were found in 6 and 4 studies, respectively. According to Table~\ref{tab:synthesis_5}, the existing distributed \gls{vm} consolidation approaches optimize a total of twelve different objectives, which can be classified into four higher-level categorizes, namely network, performance, energy, and cost. However, there is a clear emphasis on minimizing energy consumption.

\begin{table}[!ht]
\centering
\small
\caption{Names of optimization objectives}
\label{tab:synthesis_5}
\begin{tabular}{|l|p{6.1cm}|p{5.4cm}|c|}
\hline
\textbf{Category} &\textbf{Name of objective} & \textbf{Study \glsentrytext{id}s} & \textbf{Count} \\ \hline

\multirow{3}{*}{Network} & Minimizing migration cost	& S1, S15, S16 &	3 \\ \cline{2-4}
 & Minimizing network communication cost & S1, S17, S18	& 3 \\ \cline{2-4}
 & Minimizing the number of \glsentrytext{vm} migrations &	S3, S4, S5, S6 &	4 \\ \hline

\multirow{4}{*}{Performance} & Minimizing the risk of resource contention & S1	& 1 \\ \cline{2-4}
 & Minimizing \glsentrytext{sla} violations	& S2, S3, S8, S9, S15, S18	& 6 \\ \cline{2-4}
 & Minimizing delay cost & S10 & 	1 \\ \cline{2-4}
 & Maximizing load balancing	& S1, S11, S16	& 3 \\ \hline

\multirow{3}{*}{Energy} & Minimizing energy consumption & S3, S6, S8, S9, S10, S11, S12, S13, S14, S15, S18 & 11 \\ \cline{2-4}
 & Maximizing the number of released \glspl{pm} &	S4, S7	& 2 \\ \cline{2-4}
 & Maximizing resource efficiency &	S1, S5	& 2 \\ \hline

\multirow{2}{*}{Cost} & Minimizing the overall operational cost & 	S15, S19 &	2 \\ \cline{2-4}
 & Minimizing rental cost &	S2	& 1 \\ \hline

\end{tabular}
\end{table}

\begin{figure}[!ht]
	\centering	
    \includegraphics[trim=40 60 60 40, clip, width=0.89\textwidth]{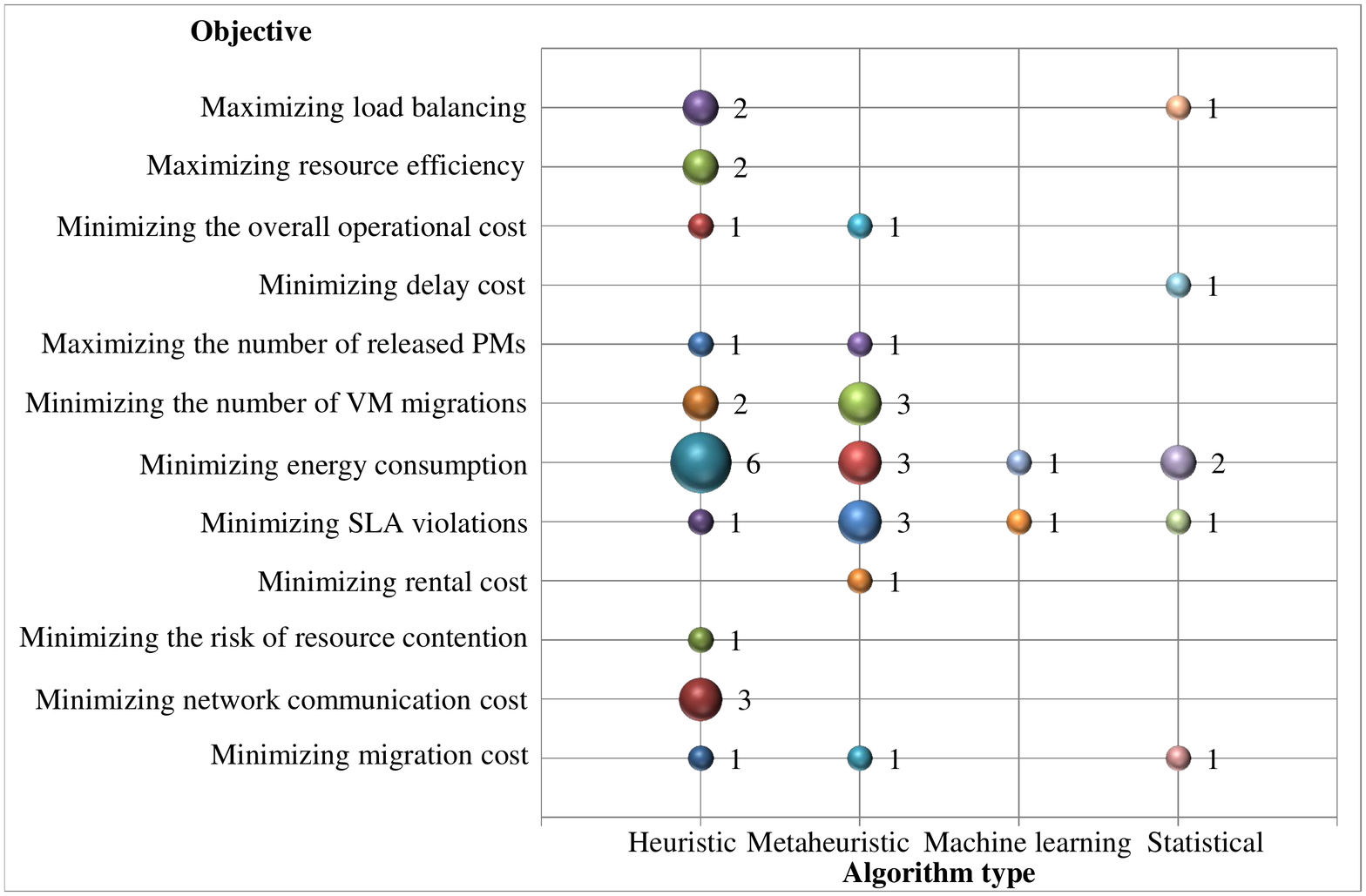}
	\caption{Number of primary studies (as bubble size) with respect to algorithm types and optimization objectives}
	\label{fig:algo_types_objective_names}
\end{figure}

Figure~\ref{fig:algo_types_objective_names} presents the number of primary studies (as bubble size) with respect to the four different algorithm types and the twelve different optimization objectives. Once again, S6 appears twice under minimizing the number of \gls{vm} migrations and under minimizing energy consumption. The results in Figure~\ref{fig:algo_types_objective_names} show that minimizing energy consumption and minimizing \gls{sla} violations are the two common objectives in all four algorithm types. The bubble plot in Figure~\ref{fig:algo_types_objective_names} also highlights a number of research gaps for potential future primary studies. It shows that 9 out of 12 optimization objectives are currently being addressed with only one or two algorithm types. For example, minimizing network communication cost appears only in heuristic approaches. So far, the other three algorithm types have not been investigated for this objective. Hence, there is an opportunity to develop new distributed \gls{vm} consolidation approaches to bridge these research gaps.

\subsection{RQ4: Evaluation Methods and Tools}
We also extracted data to categorize the primary studies on the basis of the evaluation methods used. We found that experiment is the most common evaluation method for distributed \gls{vm} consolidation approaches as it was used in all 19 primary studies. Moreover, analytical models were used in 4 studies to additionally provide a proof of a theoretical approximation or limit. Therefore, experiment is the main evaluation method.

Table~\ref{tab:synthesis_8} categorizes the primary studies on the basis of the different types of evaluation tools and load patterns used in the experiments. The results show that the most common evaluation tool in distributed \gls{vm} consolidation approaches is simulation as it was used in 16 studies. Moreover, 3 studies presented a prototype implementation. Therefore, simulation is currently the most popular evaluation tool for distributed \gls{vm} consolidation approaches. According to Table~\ref{tab:synthesis_8}, synthetic, realistic, and hybrid load patterns were used in 11, 7 and, 3 studies, respectively. It should be noted that the experiments in S2 and S3 were based on both synthetic and realistic load patterns. The results illustrate that synthetic load patterns are the most common type of load patterns used in the existing distributed \gls{vm} consolidation approaches. Moreover, the experiments in 10 studies used simulations involving synthetic load patterns. On the other hand, there is currently only one study under each load pattern type that uses a prototype implementation. Therefore, there is clearly a need for more primary studies involving prototype implementations under all three types of load patterns.

\begin{table}[!ht]
\centering
\small
\caption{Types of evaluation tools and load patterns used}
\label{tab:synthesis_8}
\begin{tabular}{|p{2.6cm}|p{2.2cm}|p{6.3cm}|c|}
\hline
\textbf{Evaluation tool}                  & \textbf{Load pattern} & \textbf{Study \glsentrytext{id}s}                       & \textbf{Count} \\ \hline
\multirow{3}{2.6cm}{Simulation}           & Synthetic             & S1, S2, S3, S5, S7, S12, S14, S15, S16, S19             & 10        \\ \cline{2-4}
                                          & Realistic             & S2, S3, S8, S9, S10, S13                                & 6              \\ \cline{2-4}
                                          & Hybrid                & S6, S11                                                 & 2              \\ \hline
\multirow{3}{2.6cm}{Prototype}            & Synthetic             & S4                                                  & 1              \\ \cline{2-4}
                                          & Realistic             & S18                                                     & 1              \\ \cline{2-4}
                                          & Hybrid                & S17                                                     & 1              \\ \hline
\end{tabular}
\end{table}

Table~\ref{tab:synthesis_10} categorizes the primary studies based on the names of the other \gls{vm} consolidation approaches that were used in the primary studies for a comparison of the results. The results show that 9 studies did not contain a comparison of the results. Moreover, each of first-fit decreasing, static threshold approach, median absolute deviation, interquartile range, and best-fit decreasing was used in 2 studies. Each of the other 15 approaches was used in only one study. Therefore, the results show that a total of 20 different approaches were used for a comparison of the results in 10 primary studies. Moreover, most of the approaches were used in only one study. Hence, there is currently no general agreement in the cloud computing community concerning good approaches for a comparison of the results.

\begin{table}[!ht]
\centering
\small
\caption{Names of other \glsentrytext{vm} consolidation approaches used for a comparison of the results}
\label{tab:synthesis_10}
\begin{tabular}{|p{7.7cm}|p{5.9cm}|c|}
\hline
\textbf{Name of other approach} & \textbf{Study \glsentrytext{id}s} & \textbf{Count} \\ \hline

None & S5, S7, S9, S10, S13, S15, S16, S18, S19	& 9 \\ \hline
First-fit decreasing & S1, S4 &	2 \\ \hline
Centralized greedy algorithm &	S2	& 1 \\ \hline
Ant colony optimization with vector algebra &	S3 &	1 \\ \hline
Static threshold approach & S3, S8 &	2 \\ \hline
Median absolute deviation & S3, S8 &	2 \\ \hline
Interquartile range & S3, S8 &	2 \\ \hline
Local regression &	S3 & 1 \\ \hline
Sercon	& S4 &	1 \\ \hline
V-MAN	& S4 &	1 \\ \hline
Best-fit decreasing &	S6, S12	& 2 \\ \hline
Random	& S11 &	1 \\ \hline
Incremental consolidation using thresholds	& S11 &	1 \\ \hline
One-shot merge & S11 &	1 \\ \hline
Genetic algorithm &	S12 &	1 \\ \hline
Energy and migration-cost-aware approach (pMapper) & S12 &	1 \\ \hline
Probability-based approach & S12 &	1 \\ \hline
Energy-aware hierarchical scheduling & S14 &	1 \\ \hline
Energy-aware scheduling for infrastructure clouds &	S14 & 1 \\ \hline
No migration & S17 & 1 \\ \hline
Optimal &	S17 &	1 \\ \hline
\end{tabular}
\end{table}

Table~\ref{tab:synthesis_11} presents study \glspl{id} of the other distributed \gls{vm} consolidation approaches that were used for the comparison of the results in the primary studies. The results show that S4 is the only primary study in which the results of the proposed approach were compared with another distributed \gls{vm} consolidation approach (S7). It is an important result. It shows that all approaches used for the comparison of the results except V-MAN (S7) are centralized approaches. Consequently, in 9 out of 10 studies that contain a comparison of the results, the results of the proposed distributed \gls{vm} consolidation approach are compared with centralized \gls{vm} consolidation approaches. Therefore, there exists little evidence on how the different distributed \gls{vm} consolidation approaches compare to one another. For more meaningful comparisons of the results in future primary studies on distributed \gls{vm} consolidation, we recommend that one or more of the 19 approaches studied in this \gls{sms} should be considered.

\begin{table}[!ht]
\centering
\small
\caption{Study \glsentrytext{id}s of other \glsentrytext{vm} consolidation approaches used for the comparison of the results}
\label{tab:synthesis_11}
\begin{tabular}{|l|p{8.6cm}|c|}
\hline
\textbf{Study \glsentrytext{id} of other approach} & \textbf{Study \glsentrytext{id}s} & \textbf{Count} \\ \hline
None & S1, S2, S3, S5, S6, S7, S8, S9, S10, S11, S12, S13, S14, S15, S16, S17, S18, S19	& 18 \\ \hline
S7 & S4 & 1 \\ \hline
\end{tabular}
\end{table}

Another important question concerning the evaluation of a distributed \gls{vm} consolidation approach is that whether or not a statistical test was used to assess the statistical significance of the results. Therefore, we extracted this information form each primary study. The results show that none of the 19 studies includes a statistical test. Therefore, it is not a common practice to use statistical testing.

\subsection{RQ5: Publication Forums}
Table~\ref{tab:synthesis_14} presents publication forum names for the 19 distributed \glsentrytext{vm} consolidation papers over time. The results show that the papers are published in 17 different publication forums. There are only two publication forums in which more than one papers have been published. These include \gls{ieee} Transactions on Services Computing and \gls{ieee} International Conference on Cloud Computing Technology and Science. Therefore, it is difficult to conclude whether or not there are any popular publication forums for distributed \gls{vm} consolidation approaches.

\begin{table}[!ht]
\centering
\footnotesize 
\caption{Publication forums for distributed \glsentrytext{vm} consolidation papers over time}
\label{tab:synthesis_14}
\begin{tabular}{|p{13.1cm}|l|l|}
\hline
\textbf{Publication forum name} & \textbf{Year} & \textbf{\glsentrytext{id}s} \\ \hline
Future Generation Computer Systems &	2016 &	S1 \\ \hline
Proceedings of the 9th IFIP TC 6 International Conference on Networking	& 2010 & S2 \\ \hline
IEEE Transactions on Services Computing & 2015 & S3, S5 \\ \hline
IEEE 4th International Conference on Cloud Computing Technology and Science & 2012 & S4 \\ \hline
IEEE Transactions on Cloud Computing & 2015 & S6 \\ \hline
IEEE International Symposium on a World of Wireless, Mobile and Multimedia Networks & 2011 & S7 \\ \hline
IEEE 4th Symposium on Network Cloud Computing and Applications & 2015 & S8 \\ \hline
IEEE 27th International Parallel and Distributed Processing Symposium Workshops PhD Forum & 2013 & S9 \\ \hline
IEEE Communications Letters & 2016 & S10 \\ \hline
IEEE 6th International Conference on Cloud Computing Technology and Science & 2014 & S11 \\ \hline
IEEE Transactions on Systems, Man, and Cybernetics: Systems & 2016 & S12 \\ \hline
IEEE 21st International Conference on Parallel and Distributed Systems & 2015 & S13 \\ \hline
IEEE 10th International Conference on High Performance Computing and Communications \& IEEE International Conference on Embedded and Ubiquitous Computing & 2013 & S14 \\ \hline
IEEE/ACM 6th International Conference on Utility and Cloud Computing & 2013 & S15 \\ \hline
9th Asia-Pacific Symposium on Information and Telecommunication Technologies & 2012 & S16 \\ \hline
39th International Conference on Parallel Processing & 2010 & S17 \\ \hline
Computer Communications & 2014 & S18 \\ \hline
Computer Networks & 2015 & S19 \\ \hline
\end{tabular}
\end{table}

\begin{figure}[!b]
	\centering	
    \includegraphics[width=0.79\textwidth]{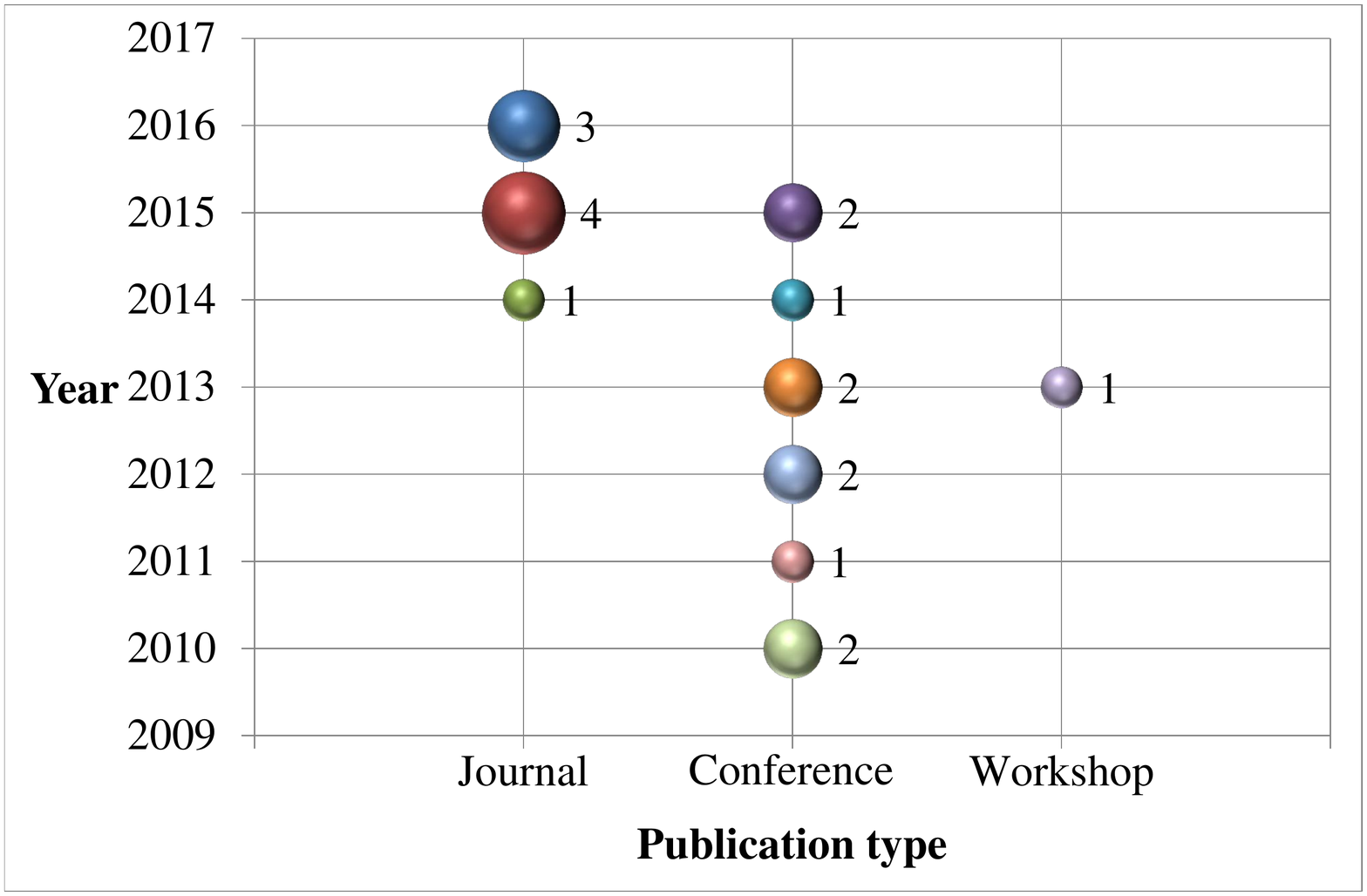}
	\caption{Number of primary studies (as bubble size) with respect to publication types over time}
	\label{fig:pub_types_years}
\end{figure}

Figure~\ref{fig:pub_types_years} presents the number of primary studies with respect to the three different publication types over time. The results show that 8 primary studies were published in journals, 10 were published in conferences, and 1 was published in a workshop. Moreover, from 2010 to 2014 (five years), a total of 10 primary studies were published. Whereas, 6 studies were published in 2015 alone. Similarly, 3 studies are published already in 2016. Furthermore, all 8 journal papers were published between 2014 and 2016. Hence, there is currently an increasing amount of interest on developing and evaluating novel distributed \gls{vm} consolidation approaches and publishing them in more rigorous publication forums.

\section{Validity Evaluation}
\label{sec:validity}
In this section, we discuss major threats to the validity of the results presented in this paper. The first main threat is related to the coverage of the relevant literature. To mitigate this threat, we designed a comprehensive search strategy based on the \gls{slr} and \gls{sms} guidelines in~\cite{Kitchenham:07, Wohlin:2012}. The search terms were extracted from the \glspl{rq} and were validated against a set of recent and prominent works on \gls{vm} consolidation including~\cite{Beloglazov:2012, Farahnakian:2015, Farahnakian2013UCC, Ferreto:2011, Hwang:2013, Marzolla:2011, Ferdaus2014, Gao:2013, Feller2012CloudCom, Farahnakian2014PDP}. Similarly, the search strings were validated against a set of known studies on distributed \gls{vm} consolidation including~\cite{Farahnakian:2015, Marzolla:2011, Feller2012CloudCom}. The search was performed in four major computer science digital libraries. Finally, the search in the \gls{acm} Digital Library was performed by using the \gls{acm} Guide to Computing Literature, which provides an expanded search.

The second threat is related to the selection of the primary studies. The results show that 75 out of 116 papers were excluded in the title and abstract level screening. It is possible that some relevant papers were erroneously excluded during the initial screening phase. To mitigate this threat, two researchers (first and second author) independently screened the titles and abstracts of each paper. The results were compared and disagreements were resolved through discussions. Moreover, for any unresolved disagreements, consensus meetings~\cite{Dyba:2008} were arranged. A similar approach was used in the full-text level screening phase.

The third major threat is related to data extraction and classification of studies. As reported by Budgen et al.~\cite{Budgen:2008}, differences in the classification of papers is a recurring theme in \glspl{sms}, even for experienced researchers. To mitigate this threat, two researchers (first and second author) independently extracted data from all 19 primary studies. The extracted data were compared and the differences were resolved in consensus meetings and by referring back to the original papers.

\section{Conclusions}
\label{sec:conclusions}
\glsresetall

In this paper, we presented a \gls{sms} of distributed \gls{vm} consolidation approaches. We used \gls{slr} and \gls{sms} guidelines in the literature to design a comprehensive search strategy. The initial search returned 202 results from four major computer science digital libraries. After the removal of duplicate results and the application of the inclusion/exclusion criteria at two levels, 21 primary studies were selected. Finally, 2 studies were excluded in the quality assessment stage, which left 19 primary studies for data extraction and synthesis.

The objective of the \gls{sms} was to provide a comprehensive, unbiased overview of the state-of-the-art on distributed \gls{vm} consolidation approaches. The \gls{sms} comprises five \glspl{rq} concerning: (1) existing approaches, (2) types of algorithms being used, (3) objectives being optimized, (4) evaluation methods and tools being used, and (5) popular publication forums over time. The results of the first \gls{rq} showed that 14 out of 19 studies presented pure distributed \gls{vm} consolidation algorithms, while 2 studies presented centralized algorithms with a distributed architecture for \gls{vm} consolidation and 3 studies presented \gls{vm} consolidation algorithms for geographically distributed data centers.

The answer to the second \gls{rq} showed that the existing distributed \gls{vm} consolidation approaches use four different types of algorithms, namely heuristics, metaheuristics, machine learning algorithms, and statistical approaches. Moreover, heuristics and metaheuristics are currently the most popular algorithm types. The most frequently used algorithm is distributed or coordinated local search heuristic, while \gls{aco} and greedy are the second most used algorithms. Only a small fraction of the existing distributed \gls{vm} consolidation approaches can be categorized as using an online optimization technique. Hence, online optimization techniques are currently not sufficiently investigated for distributed \gls{vm} consolidation.

For the third \gls{rq} concerning optimization objectives, we categorized the primary studies with respect to number and name of objectives and problem formulation. The results showed that nearly $\frac{3}{4}$ of the primary studies optimize either one or two objectives and only a few approaches optimize more than two objectives. The existing distributed \gls{vm} consolidation approaches optimize a total of 12 different objectives. The most popular optimization objective is minimizing energy consumption. Other popular objectives include minimizing \gls{sla} violations and minimizing the number of \gls{vm} migrations. 9 out of 12 objectives are currently being addressed with only one or two algorithm types. Hence, the focus of future primary studies may be directed to investigate the remaining algorithm types for optimizing these objectives. About $\frac{2}{3}$ of the studies presented a multi-objective or many-objective problem formulation with an \gls{aof}, while the rest of the studies presented a single-objective problem formulation. None of the studies presented a pure multi-objective or many-objective problem formulation. Hence, future research may be directed to develop pure multi-objective and many-objective distributed \gls{vm} consolidation approaches.

The results of the fourth \gls{rq} showed that experiment is the most common evaluation method for distributed \gls{vm} consolidation approaches. The most common evaluation tool is simulation. Moreover, synthetic load patterns are the most common type of load patterns. Therefore, simulations involving synthetic load patterns are currently the most common evaluation tool. We also extracted data with respect to the other \gls{vm} consolidation approaches that were used for a comparison of the results. The results showed that 9 studies did not contain a comparison of the results. Moreover, a total of 20 different approaches were used for a comparison of the results in the remaining 10 studies. In 9 out of 10 studies that contained a comparison of the results, the results of the proposed distributed \gls{vm} consolidation approach were compared with centralized \gls{vm} consolidation approaches. Therefore, there exists little evidence on how the different distributed \gls{vm} consolidation approaches compare to one another. Hence, there is a need for more comparative studies involving multiple distributed \gls{vm} consolidation approaches. We recommend that one or more of the 19 approaches studied in this \gls{sms} should be considered for more meaningful comparisons of the results in future studies.

The answer to the fifth \gls{rq} showed that the 19 studies were published in 17 different publication forums. Therefore, it is difficult to conclude whether or not there are any popular publication forums for distributed \gls{vm} consolidation approaches. There is currently an increasing amount of interest on developing and evaluating novel distributed \gls{vm} consolidation approaches for cloud data centers and publishing them in more rigorous publication forums.

\section*{Acknowledgements}
The authors would like to thank Muhammad Usman at Blekinge Institute of Technology for his valuable comments on the review protocol. The work was supported by the Need for Speed (N4S) Program (\url{http://www.n4s.fi}). Adnan Ashraf also received a research grant from the Ulla Tuominen Foundation.

\bibliographystyle{abbrv}
\bibliography{references}

\end{document}

%% file: glossary.tex
\newacronym{tivit}{TiViT}{Tieto- ja Viestint\"ateollisuuden Tutkimus}
\newacronym{tekes}{Tekes}{Teknologian ja innovaatioiden kehitt\"amiskeskus}
\newacronym{saas}{SaaS}{Software as a Service}
\newacronym{paas}{PaaS}{Platform as a Service}
\newacronym{iaas}{IaaS}{Infrastructure as a Service}
\newacronym{qos}{QoS}{Quality of Service}
\newacronym{ec2}{EC2}{Elastic Compute Cloud}
\newacronym{vm}{VM}{Virtual Machine}
\newacronym{rtt}{RTT}{Round Trip Time}
\newacronym{cpu}{CPU}{Central Processing Unit}
\newacronym{cgi}{CGI}{Common Gateway Interface}
\newacronym{tcp}{TCP}{Transmission Control Protocol}
\newacronym{http}{HTTP}{Hypertext Transfer Protocol}
\newacronym{osgi}{OSGi}{Open Services Gateway initiative}
\newacronym{uri}{URI}{Uniform Resource Identifier}
\newacronym{url}{URL}{Uniform Resource Locator}
\newacronym{aws}{AWS}{Amazon Web Services}
\newacronym{ami}{AMI}{Amazon Machine Image}
\newacronym{jvm}{JVM}{Java Virtual Machine}
\newacronym{ssl}{SSL}{Secure Socket Layer}
\newacronym{osi}{OSI}{Open Systems Interconnection}
\newacronym{acl}{ACL}{Access Control List}
\newacronym{ria}{RIA}{Rich Internet Application}
\newacronym{api}{API}{Application Programming Interface}
\newacronym{rmi}{RMI}{Remote Method Invocation}
\newacronym{rpc}{RPC}{Remote Procedure Call}
\newacronym{pd}{PD}{Proportional-Derivative}
\newacronym{pi}{PI}{Proportional-Integral}
\newacronym{nist}{NIST}{National Institute of Standards and Technology}
\newacronym{sbac}{SBAC}{Session-Based Admission Control}
\newacronym{cbac}{CBAC}{Cost-Based Admission Control}
\newacronym{sla}{SLA}{Service Level Agreement}
\newacronym{rmse}{RMSE}{Root Mean Square Error}
\newacronym{nrmse}{NRMSE}{Normalized Root Mean Square Error}
\newacronym{ema}{EMA}{Exponential Moving Average}
\newacronym{ssh}{SSH}{Secure Shell}
\newacronym{rest}{REST}{Representational State Transfer}
\newacronym{war}{WAR}{Web Application Archive}
\newacronym{ip}{IP}{Internet Protocol}
\newacronym{aces}{ACES}{Admission Control based on Estimation of Service times}
\newacronym{cosac}{CoSAC}{Coordinated Session-based Admission Control}
\newacronym{cramp}{CRAMP}{Cost-efficient Resource Allocation for Multiple web applications with Proactive scaling}
\newacronym{acvas}{ACVAS}{adaptive Admission Control for Virtualized Application Servers}

\newacronym{it}{IT}{Information Technology}
\newacronym{aco}{ACO}{Ant Colony Optimization}
\newacronym{rq}{RQ}{Research Question}
\newacronym{gop}{GOP}{Group of Pictures}
\newacronym{sbacs}{SBACS}{Stream-Based Admission Control and Scheduling}
\newacronym{gae}{GAE}{Google App Engine}
\newacronym{pso}{PSO}{Particle Swarm Optimization}
\newacronym{s3}{S3}{Simple Storage Service}
\newacronym{ctt-sp}{CTT-SP}{Cost Transitive Tournament Shortest Path}
\newacronym{ddg}{DDG}{Data Dependency Graph}
\newacronym{pm}{PM}{Physical Machine}
\newacronym{mmas}{MMAS}{Max-Min Ant System}
\newacronym{acs}{ACS}{Ant Colony System}
\newacronym{as}{AS}{Ant System}
\newacronym{tsp}{TSP}{Travelling Salesman Problem}
\newacronym{sbse}{SBSE}{Search-Based Software Engineering}

\newacronym{acm}{ACM}{Association for Computing Machinery}
\newacronym{ieee}{IEEE}{Institute of Electrical and Electronics Engineers}
\newacronym{picoc}{PICOC}{Population, Intervention, Comparison, Outcomes, and Context}
\newacronym{slr}{SLR}{Systematic Literature Review}
\newacronym{sms}{SMS}{Systematic Mapping Study}
\newacronym{id}{ID}{identifier}
\newacronym{aof}{AOF}{Aggregate Objective Function}
